\documentclass{article}
\usepackage{amssymb}

\newcommand{\reals}{{\mathbb R}}

\begin{document}

\title{Comment on ``On the proper behavior of atoms'' by Paul Anglin}
\author{Russell K. Standish \\
Mathematics and Statistics, University of New South Wales \and Stephen L.
Keen \\
Economics and Finance, University of Western Sydney }
\maketitle

\begin{abstract}
  Paul Anglin criticised our analysis of the neoclassical theory of
  the firm, but makes a number of incorrect assertions about our
  assumptions. We correct these misunderstandings, but acknowledge
  that one criticism he makes is correct. We correct this flaw with a
  new argument that supersedes the flawed strategic reaction argument
  we presented in our previous paper.
\end{abstract}

\section{The profit formula}

We take as our starting point, the usual profit formula of a single
product market with $n$ firms:
\begin{equation}\label{profit}
\pi_i = q_iP(Q)-\int_0^{q_i} MC(q_i) dq_i,
\end{equation}
where $\pi_i$ is the profit obtained by firm $i$, as a function of its
production $q_i$, and the total market production $Q=\sum_i q_i$. The
function $P(Q)$ is the demand curve, namely the price the good
achieves when $Q$ items of the good is available on the market. The
function $MC(q_i)$ is the marginal cost of producing an extra item of the
good, given that a firm is producing $q_i$ items.

\section{The trouble with derivatives}

In \cite{Anglin08}, Paul Anglin critiques our paper \cite{Keen-Standish06}.
We note a number of problems with this critique.

Anglin's initial proposition is that our results depend on the size of the
increment to output for each firm:

\begin{quotation}
I suggest that a flawed premise is being used since it is also true that the
effect on P would be about 100 times larger if the change in output by a
single firm increased from $dq_i$ = 1 to 100. So, before analyzing the effect
of a change by a mass of firm, a more relevant question is: is $dq_i$ = 1 or
100 (or $-1$ or $-100$)?\cite[p. 278]{Anglin08}
\end{quotation}

However, our argument was based not on discrete changes to output but on
derivatives. The $dq_{i}$ he mentions is \textquotedblleft
infinitesimal\textquotedblright : it cannot be equal to 1 or 100. In any
case, we do not use \textquotedblleft infinitesimals\textquotedblright ,
which are mathematically problematic, but regular derivatives, which in the
case of a multivariate function $y(q_{1},\ldots ,q_{n})$ can be either
partial $\partial y/\partial q_{i}$ or total $dy/dx$.

In footnote 1, Anglin conjectures that the relation $dq_{i}/dQ=\sum_{j}%
\partial q_{i}/\partial q_{j}$ \textquotedblleft seems to be a consequence
of the fact that $Q=\sum_{j}q_{j}$\textquotedblright\ \cite[p. 278]{Anglin08}%
. In comments he made on a previous version of this paper, it would
appear that this is the crux of his disagreement with our analysis. In
\cite{Keen-Standish06}, we effectively assumed that
\begin{equation}
\frac{dq_i}{dQ}=\sum_j\frac{\partial q_i}{\partial q_j}=1
\end{equation}
in going from equation (4) to (6) in that paper. On reflection, we
realise this criticism is correct --- there is no justification for
assuming $dq_i/dQ$ has any particular value. Nevertheless, the Keen
result (eq 6 of \cite{Keen-Standish06}) can still be derived as the
system equilibrium assuming a much weaker additional condition that
$dq_i/dQ = dq_j/dQ, \forall i,j$ holds at equilibrium.

\section{Symmetry of firms}

In footnote 2 of \cite{Anglin08}, Anglin asserts we made a symmetry
assumption $Q=nq_{i}$, from which he derives an inconsistency. We did
not make this assumption at any point in our paper. In the referees
comments he made on an earlier version of this paper, it would appear
that this is a derived consequence of our assumption that $dq_i/dQ=1$.
Coupled with the boundary condition $Q=0 \Rightarrow q_i=0$ and
integrating, this would imply $q_i=Q/n$.

However, since the Keen equilibrium only requires that $dq_i/dQ=dq_j/dQ,
\forall i,j$ at equilibrium, there is no specific requirement for the
market to be evenly shared amongst the firms, except in the case of
constant marginal cost, as detailed in section \ref{inhomog cv}.

We do assume that each firm has identical marginal cost functions
$MC(q_i)$, which is also assumed in the traditional presentation of
the Cournot profit maximum. This is for pedagogical convenience
however, the argument presented in section \ref{inhomog cv} does not
depend on this assumption, and can be easily generalised.

\section{Total derivative with respect to industry output rather than single
firm's output}

The traditional analysis of the Marshallian and Cournot models is to
to hypothesize behavior by the individual firm such that it sets the
partial derivative $\partial\pi_i/\partial q_i = 0$ (see e.g.
\cite[eq (2)]{Keen-Standish06}); in the Marshallian model this is
described as \textquotedblleft atomistic\textquotedblright\ 
profit-maximizing behavior, while in the Cournot model it is described
as a constrained profit level in response to the strategic responses
of other firms. The Marshallian proposition is strictly false, since
the profit of a single firm $\pi_i$ is a function of all $n$ firms'
outputs $q_i$, not a single variable function, whether or not the
individual firm can in fact affect the behavior of other firms. The
extrema of an $n$-variable function is found at the zero of the
derivative, ie when all partial derivatives $\partial\pi_i/\partial
q_j = 0$. However
\begin{equation}
\partial\pi_i/\partial q_j = \delta_{ij}(P-MC) + q_iP'
\end{equation}
which can never be satisfied where $q_i>0$ and $P'<0$. The condition
$\partial\pi_i/\partial q_i = 0$ describes an unstable equilibrium ---
it is vulnerable to firms pulling in the same direction, which can
happen even in the absence of explicit collusion \cite{Standish-Keen04}.

Instead we propose the condition that all firm's profits are maximised
with respect to total industry output $d\pi_i/dQ=0$. This constrains
the dynamics of firms' outputs to an $n-1$-dimensional polyhedron, but
otherwise does not specify what the individual firms should do. As an
equilibrium condition, it is vulnerable to a single firm ``stealing''
market share. However, no firm acts in isolation. The other firms will
react, negating the benefit obtained by first firm, causing the system
to settle back to the  $d\pi_i/dQ=0$ manifold.

\section{Conjectural Variation}

In our paper, we introduce the idea of firms reacting to the
production decisions of their competitors by introducing a dependence
between our previous independent variables $q_{i}$. We thank Anglin
for reminding us of considerable previous history of doing this under
the name of ``conjectural variation''; however, this was a literature
of which we were already aware, and whose use of the
concept differs from our purpose in introducing it here.\footnote{%
  This can be interpreted as firms anticipating what their competitors
  might do, although we tend to regard it as describing reactions to
  competitors in a ``time-free'' model, so the variation is not
  conjectural but reactionary.} Our intent was to make a mathematical
argument that shows what happens in the Cournot analysis when one
relaxes the assumption of atomism. We have not attempted to model any
form of conjectural variation or reaction by the firms in the agent
model, and in any case the agent model does not have the atomistic
constraint imposed upon it.

We appreciate the reference \cite{Kamien-Schwartz83} Anglin provided, but
note that as they started from the incorrect differential condition ($%
\partial\pi_i/\partial q_i=0$), their results are not applicable.

In the next section, we present a strategic response argument that
does not make use of the conjectural variation idea at all.

\section{Evolution of $dq_i/dQ$}

\label{inhomog cv}

In our paper \cite{Keen-Standish06}, we introduced a homogenous conjectural
variation parameter $\partial q_i/\partial q_j = \theta$. As pointed
out by Anglin, this analysis makes use of the faulty assumption
$dq_i/dQ=\sum_j \partial q_i/\partial q_j$. To circumvent this
problem, and generalise the argument, we take the point that $dq_i/dQ$
are unconstrained endogenous variables, and so we introduce the variables 
\begin{equation}
\frac{dq_i}{dQ} = \theta_i.
\end{equation}
This extends phase space from the $n$-dimensional space of firm
production $q_i, i=1\ldots n$ to a $2n-1$-dimensional phase space,
with the constraint
\begin{equation}
\sum_i\frac{dq_i}{dQ} = \frac{dQ}{dQ} =1.
\end{equation}
The $\theta_i$ might be thought of as a firm's response function to
changing industry output.

With the usual profit formula (\ref{profit}) the maximum profit for a
single firm obtains at the zero of
\begin{eqnarray}  \label{max-profit}
\frac{d\pi_i}{dQ} &=& P \frac{dq_i}{dQ} + q_i \frac{dP}{dQ} -
MC(q_i)\frac{dq_i}{dQ}  \nonumber \\
&=& P\theta_i + q_iP^{\prime} - MC(q_i)\theta_i
\end{eqnarray}

We may sum equation (\ref{max-profit}) over $i$ to obtain
\begin{eqnarray}  \label{max-industry-profit}
P + QP^{\prime} - \sum_i MC(q_i)\theta_i = 0.
\end{eqnarray}

Given a fixed market partition $\{s_i=q_i/Q\}$, the maximum profit obtains
at the zero of the derivative of the total industry profit
\begin{equation}\label{max-profit-given-partition}
\frac{d}{dQ}\left(QP-\sum_i\int^{s_iQ}MC(q)dq\right) = P + QP' - \sum_i s_iMC(q_i) = 0.
\end{equation}
Comparing equations (\ref{max-industry-profit}) and
(\ref{max-profit-given-partition}), we see that the individual firm
profit is submaximal unless
\begin{equation}\label{profit-cond}
 \sum_i MC(q_i)(s_i-\theta_i) = 0.
\end{equation}
The vector $(m_i=MC(q_i))$ lies in the positive cone $\reals^{n+}$
(ie $m_i>0,\, \forall i$). The vector $(t_i=s_i-\theta_i)$ lies on a
hyperplane passing through the origin, and perpendicular to the unit vector
$(1,1,...1)$, since $\sum_i t_i=0$. Condition (\ref{profit-cond})
can be thought of as a dot product $\mathbf{m}\cdot {\bf t}=0$. This condition can
only be satisfied if $\mathbf{m}$ is proportional to the unit vector (ie
marginal cost is constant) or ${\bf t}=0$, which implies
$\theta_i=s_i,\,\forall i$.  Given a particular partition of the
market, profit of all firms will always be increased by moving the
$\theta_i$ variables closer to the market share $s_i$.

Substituting this condition for variable marginal cost into (\ref{max-profit}) gives:
\begin{equation}\label{s_i}
s_iP+s_iQP'-s_iMC(s_iQ)=0
\end{equation}
which can only be simultaneously satisfied for all $i$ if the market is
equipartitioned ($s_i=1/n$).

The Keen equilibrium obtains on the manifold where
$\theta_i=1/n$. Substituting this into equation (\ref{max-profit}),
one obtains 
\begin{equation}\label{Keen}
P-MC(q_i) + nq_iP' = 0
\end{equation}
which can be rearranged to yield
\begin{equation}
MR_i-MC = P(Q)+q_iP'(Q)  -MC(q_{i})  =\frac{n-1}{n}(
P-MC(  q_{i})) \label{MR_MCGapRule}%
\end{equation}
where $MR_i$ is the marginal revenue of the firm.

When marginal cost is constant, equation (\ref{max-industry-profit})
implies that the industry operates at the monopoly pricing at
equilibrium:
\begin{equation}
P+QP' -MC = 0
\end{equation}
and from (\ref{max-profit}) we see
\begin{equation}
q_i=\theta_iQ
\end{equation}
Only when $\theta_i=1/n$ does this coincide with the Keen equilibrium.

We may rearrange equation (\ref{Keen}) to give
\begin{equation}\label{q_i}
q_i=\frac{MC(q_i)-P}{nP'}
\end{equation}
If the right hand side of this equation were a monotonic decreasing
function of $q_i,\, \forall q_j, j\ne i$, then a unique solution
exists for $q_i$, the market is equipartitioned between firms and the
Keen equilibrium coincides with monopoly pricing. However, if multiple
solutions to (\ref{q_i}) exist,\footnote{ For example with
  $P(Q)=10-Q$ and $MC(q)=1/q$, $(P-MC)/P'$ exhibits a peak value at an
  intermediate value, so is not monotonic. We thank Paul Anglin for
  providing this example.  } 
then the market need not be
equipartitioned, and in general the Keen equilibrium differs from
monopoly pricing. However, in the limit $n\rightarrow\infty$, assuming
finite total industry output, $q_i$ is $o(1/n)$, so $P-MC(q_i)$ tends
to some positive value, differing from competitive pricing.

In the simple case of a linear demand curve, multiple solutions to
$q_i$ can only exist for falling marginal cost. Such markets are
dominated by a scramble for market share, as there is a distinct
``economy of scale'' advantage to being market leader. The analysis
presented here does not help determine what the equilibrium state will be.

If the marginal cost function differed between firms, the result
$\theta_i=s_i$ still holds. The main difference is that the
corresponding equation (\ref{s_i}) is now firm dependent
\begin{equation}\label{diffMC_i}
P+QP'-MC_i(s_iQ)=0,
\end{equation}
and the market is no longer equipartitioned at equilibrium. The
equivalent of (\ref{Keen}) is
\begin{equation}
MR_i - MC_i = \frac{Q-q_i}{Q}(P-MC_i(q_i)).
\end{equation}

\section{Agent simulation}

What evidence is there that the parameters $\theta_i$ introduced in
the previous section will undergo evolution so as to optimise the
profit levels of the firms? In \cite{Standish-Keen04}, we introduced a
simple agent based model which exhibited an interesting emergent
phenomenon where agents would lock into the same strategy of
decreasing production to improve profits. At the start of the
simulation, agents are randomly increasing or decreasing their
production levels without affecting total industry production much. In
terms of $\theta_i$, this implies $|\theta_i|\gg 1/n$, and total
industry production from equation (\ref{max-profit}) is close to
competitive levels. As the emergent lock in effect takes place, the
firms are changing their production levels in the same way, so
$\theta_i=1/n$, and the system converges to the Keen equilibrium. 

Qualitatively, the results of the two models do differ, with the agent
model exhibiting a range of convergent behaviour not seen in the
differential case. In the agent model, we were also able to reproduce
the neoclassical result of convergence by the firms to output levels
at which each firm's marginal cost equaled its marginal revenue in two
ways. However, neither of these accord with the standard
\textquotedblleft Marshallian\textquotedblright\ or \textquotedblleft
Cournot\textquotedblright\ explanations. Convergence to the Cournot
output level occurred:

\begin{enumerate}
\item When a fraction of firms behaved irrationally, by continuing with a
strategy (for example, increasing output) when that strategy \emph{reduced}
profit in the previous iteration. Convergence to the neoclassical
expectation was monotonic as the proportion of irrational firms was raised
from zero to 25 percent; from 25 to 50 percent, the neoclassical case
applied; while above 50 percent irrational behaviour, the firms and the
system followed a random walk. This result was independent of the number of
firms in the industry; and

\item As the standard deviation $\sigma $ of the parameter $\delta q_{i}$
rose, as shown in our paper. This result was also independent of the number
of firms in the industry.
\end{enumerate}

\section{Conclusion}

Our conclusions about the strict falsity of the Marshallian model, and the
lack of content of the Cournot model---in that while it is strictly true,
actual profit-maximizers would not play the Cournot-Nash game---still stand.
We therefore continue to assert that economics does not have a model of
price setting. Blinder et al.\cite{Blinder-etal98}, provides a good empirical
survey of price setting practices in the real world, and as with our model,
this survey strongly contradicts accepted neoclassical beliefs. We suggest
that a good research goal for economists would be to devise a model of
competition that replicates the results of this study.


\end{document}